# Diabetes detection using deep learning techniques with oversampling and feature augmentation

María Teresa García-Ordás[a], Carmen Benavides[b], José Alberto Benítez-Andrades[b,*], Héctor Alaiz-Moretón[a], Isaías García-Rodríguez[a]

[a] *SECOMUCI Research Groups, Escuela de Ingenierías Industrial e Informática, Universidad de León, Campus de Vegazana s/n, León C.P. 24071, Spain*
[b] *SALBIS Research Group, Department of Electric, Systems and Automatics Engineering, Universidad de León, Campus of Vegazana s/n, León, León, 24071, Spain*

**ABSTRACT**

*Background and objective*: Diabetes is a chronic pathology which is affecting more and more people over the years. It gives rise to a large number of deaths each year. Furthermore, many people living with the disease do not realize the seriousness of their health status early enough. Late diagnosis brings about numerous health problems and a large number of deaths each year so the development of methods for the early diagnosis of this pathology is essential.

*Methods*: In this paper, a pipeline based on deep learning techniques is proposed to predict diabetic people. It includes data augmentation using a variational autoencoder (VAE), feature augmentation using an sparse autoencoder (SAE) and a convolutional neural network for classification. Pima Indians Diabetes Database, which takes into account information on the patients such as the number of pregnancies, glucose or insulin level, blood pressure or age, has been evaluated.

*Results*: A 92.31% of accuracy was obtained when CNN classifier is trained jointly the SAE for featuring augmentation over a well balanced dataset. This means an increment of 3.17% of accuracy with respect the state-of-the-art.

*Conclusions*: Using a full deep learning pipeline for data preprocessing and classification has demonstrate to be very promising in the diabetes detection field outperforming the state-of-the-art proposals.

## 1. Introduction

Diabetes is a chronic pathology that occurs when the amount of glucose in blood is too high. Glucose is the body's main source of energy and insulin is the hormone, secreted by the pancreas that regulates the amount of glucose in the cells to be used for energy. Diabetic people do not produce enough insulin so the glucose remains in the blood [15].

Having too much glucose in blood, may cause a number of health problems [21], such as heart and dental diseases, stroke, eye problems, nerve damage, etc.

In 2020, Olawsky et al. [22] carried out an study to evaluate the relationship of glycemic variability and 5 year hypoglycemia risk in patients with Type 2 Diabetes (T2DM).

Diabetes gives rise to a large number of deaths each year. Furthermore, a lot of people that live with the disease do not realize the seriousness of their health condition early enough. The number of diabetic people is predicted to increase year by year [17]. Many complications occur if diabetes remains untreated and unidentified. In order to reduce the number of deaths brought about by diabetes, the development of methods and techniques for the early diagnosis of diabetes is essential, as a large number of deaths in diabetic patients are due to a late diagnosis.

A number of techniques over the years have been developed to deal with the detection problem. In the work developed by Sisodia et al. [27], three machine learning classification techniques were used: decision tree, support vector machine (SVM) and naive Bayes. In these cases, the naive Bayes algorithm outperforms the other classifiers by obtaining an accuracy of 76.30%.

In [18], diabetes is predicted using significant attributes, with the relationship between the differing attributes also being characterized. The selection of significant attributes was made using the principal component analysis method. The authors found a strong relationship between diabetes and body mass index (BMI) and with the glucose level. After this process, artificial neural network (ANN), random forest (RF) and k-means clustering techniques were implemented for the classification step, with the best accuracy obtained using the ANN technique with a 75.7% of success rate.

* Corresponding author.
  *E-mail addresses:* mgaro@unileon.es (M.T. García-Ordás), mcbenc@unileon.es (C. Benavides), jbena@unileon.es (J.A. Benítez-Andrades), hector.moreton@unileon.es (H. Alaiz-Moretón), isaias.garcia@unileon.es (I. García-Rodríguez).

The use of a rule extraction algorithm, Re-RX with J48graft, combined with sampling selection techniques (sampling Re-RX with J48graft) was proposed by Hayashi et al. [8] with the same purpose, obtaining results of up to 83.83%.

Fuzzy classification rules are more interpretable with respect to other rules. For this reason, fuzzy classification rules are used extensively in the classification and decision support systems for disease diagnosis: Polat et al. [23], use principal component analysis (PCA) and an adaptive neuro-fuzzy inference system (ANFIS) with the combination of both methods achieving a performance of 89.47%. Similar results are obtained in Lukmanto et al. [17] in which feature selection is used to identify the valuable features in the dataset. SVM model is then train with the selected features to generate the fuzzy rules and fuzzy inference process is finally used to classify the output.

Siva et al. [5] present a diabetes prediction model using the concept of fuzzy rule and grey wolf optimization but the results are not very promising in comparison with other methods in the literature. Fuzzy system is also employed in Mansourypoor and Asadi [20]. In this study, a reinforcement learning-based evolutionary fuzzy rule-based system (RLEFRBS) is developed for diabetes diagnosis. The authors tested their method in two datasets obtaining results of up to 84% in Pima-Indian diabetes dataset. In [3], a hybrid decision support system based on rough set theory (RST) and bat optimization algorithm (BA) called RST-BatMiner is presented. In the first step, the data is preprocessed and redundant features are removed. In the second stage, for each class BA is invoked to generate fuzzy rules by minimizing proposed fitness function. Finally, an ada-boosting technique is applied to the rules generated by BA to increase the accuracy rate of generated fuzzy rules. They achieved a performance of 85.33% in the Pima-Indian dataset.

Alneamy et al. [19], have developed a method based on the Teaching Learning-Based Optimization (TLBO) algorithm and Fuzzy Wavelet Neural Network (FWNN) with Functional Link Neural Network (FLNN). They tested the efficiency of their method in five different datasets for different purposes, obtaining a 88.67% of accuracy for the Pima-Indian diabetes dataset, which is a very promising result.

More machine learning techniques other than fuzzy methods are also widely used to try to deal with this problem. In [10] the authors decided to use machine learning techniques to solve the problem, achieving a 86.26% success rate. They extracted the features from the dataset using stacked autoencoders and the dataset is classified using a softmax layer. Furthermore, the fine tuning of the network is done using backpropagation with the training dataset.

In [12] five different predictive models were tested: Linear Kernel Support Vector Machine (SVM-linear), Radial Basis Function (RBF) Kernel Support Vector Machine, K-Nearest Neighbour (k-NN), Artificial Neural Network (ANN) and Multifactor Dimensionality Reduction (MDR) obtaining the best results, with a success rate of 89% using SVM.

Singh et al. [26], developed a stacking-based evolutionary ensemble learning system called NSGA-II-Stacking. They carried out a data pre-processing step, filling outliers and missing values with the median values. For base learner selection, a multi-objective optimization algorithm is used. As for model combination, k-nearest neighbor is employed as a meta-classifier that combines the predictions of the base learners. The comparative results demonstrate that their proposal achieves an accuracy of 83.8%.

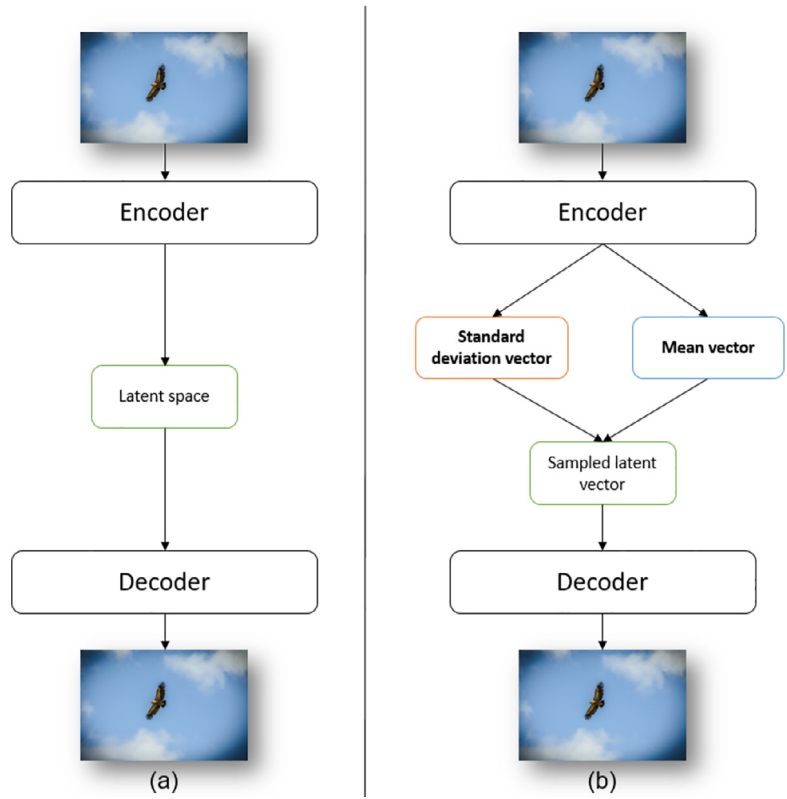

**Fig. 1.** In (a), the vanilla autoencoder with a simple latent vector. In (b) a variational autoencoder scheme with the mean and standard deviation layers used to sample the latent vector.

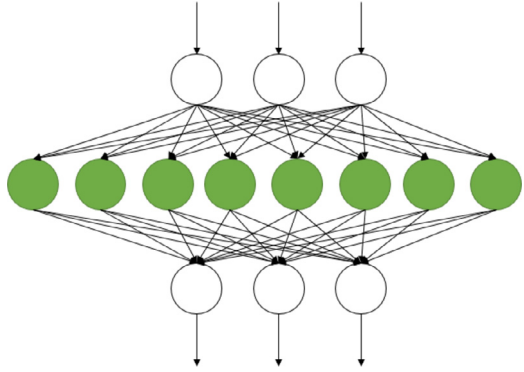

**Fig. 2.** Example of an SAE network architecture. In the latent space there are more neurons than in the input and output and L1 regularization term is applied.

All of these works, were developed on the Pima-Indian dataset, which has been a challenge for many years.

Deep learning has been demonstrated to be able to solve many complex problems in recent years. Stacked autoencoders (SAE) [16,31,32], Deep Belief Networks (DBN) [28], Long Short Term Memory (LSTM) [30] or Convolutional Neural Networks (CNN) [25,29] are some examples of successful deep learning solution proposed in the last year.

In this paper, a fully deep learning pipeline for diabetes prediction is proposed. Variational autoencoder (VAE), Sparse autoencoder (SAE) and Convolutional neural network (CNN) are existing technologies but rarely are used all together. In this work, we have carried out data augmentation both in samples (VAE) and features (SAE). SAE has been trained jointly with a CNN classifier which allow them to get feedback from each other in the backpropagation step improving the quality of the features generated according to the spatial representation forced by CNN.

The paper is organised as follows: In Section 2, the data preprocessing steps and the different techniques used to carry out the feature augmentation are detailed. Vanilla multilayer perception and convolutional neural networks are also introduced. The experiments, results with all of the techniques are shown, discussed and compared with other state-of-the-art works in Section 3 and finally, we conclude in Section 4.

## 2. Methodology

### 2.1. Data preprocessing: normalization

When a dataset is used to train a model, every of their features usually follows a different distribution. In these cases, it is very difficult for an artificial neural network to fit the data. To solve this problem, there are so many different techniques which try to adjust every feature to obtain a similar range in the real numbers set. Some of the most typical normalizers are:

- MaxMin Normalization takes into account the maximum and the minimum values to fix the data to into the range [0,1] following Eq. 1.

$$\hat{x} = \frac{x - x_{\min}}{x_{\max} - x_{\min}} \quad (1)$$

where $x$ is the sample, $x_{\min}$ is the minimum value and $x_{\max}$ which is the maximum value of each feature.

- Standard Normalization uses the statistical information of the distribution to adjust the data in order to have mean equal to 0 and a standard deviation of 1. See Eq. (2).

$$\hat{x} = \frac{x - \mu}{\sigma} \quad (2)$$

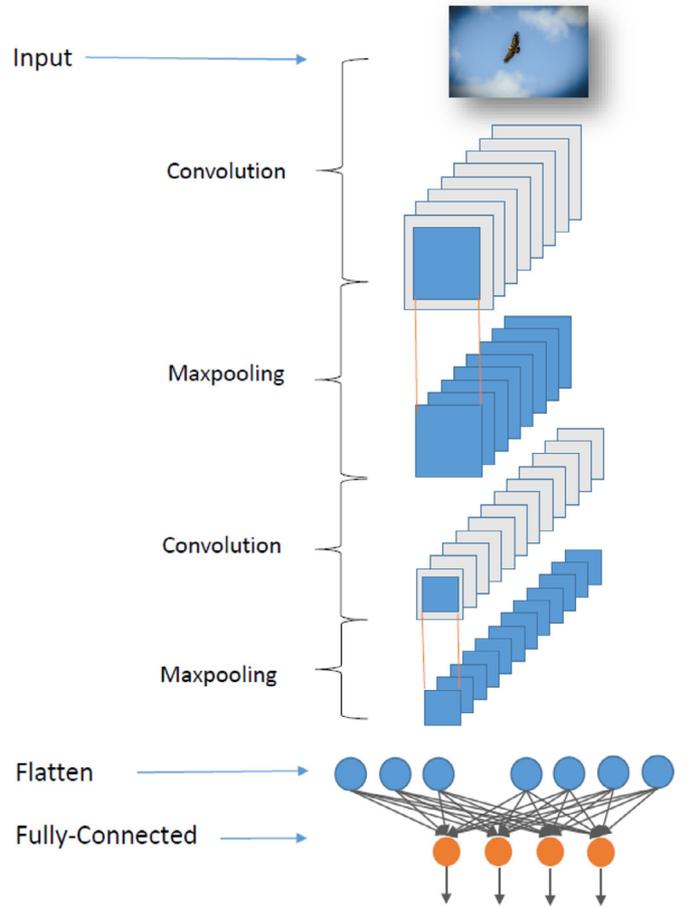

**Fig. 3.** A vanilla CNN representation over an example image.

with mean $\mu$:

$$\mu = \frac{1}{N} \sum_{i=1}^{N} x_i \quad (3)$$

and standard deviation $\sigma$:

$$\sigma = \sqrt{\frac{1}{N} \sum_{i=1}^{N} (x_i - \mu)^2} \quad (4)$$

where N is the number of samples and $x_i$ is the $i$th element of the dataset.

- Logarithmic Normalization applies the logarithmic scale to the data following Eq. (5).

$$\hat{x} = \log(x) \quad (5)$$

### 2.2. Data augmentation: variational autoencoder (VAE)

Frequently, when processing a labeled dataset it can be seen that some of the classes predominated over all the others. This is very common in medical datasets where the percentage of some rare diseases samples used to be lower than the healthy ones. This can cause that the machine learning techniques does not focus on these small classes and learns by only taking the most crowded class into account. In order to solve this, there are two main research lines: undersampling and oversampling. In undersampling techniques, the goal is to reduce the most representative class so that all of them have a similar number of elements. On the other hand, oversampling methods aims to increase the number of elements in the less representative classes. The Synthetic Minority

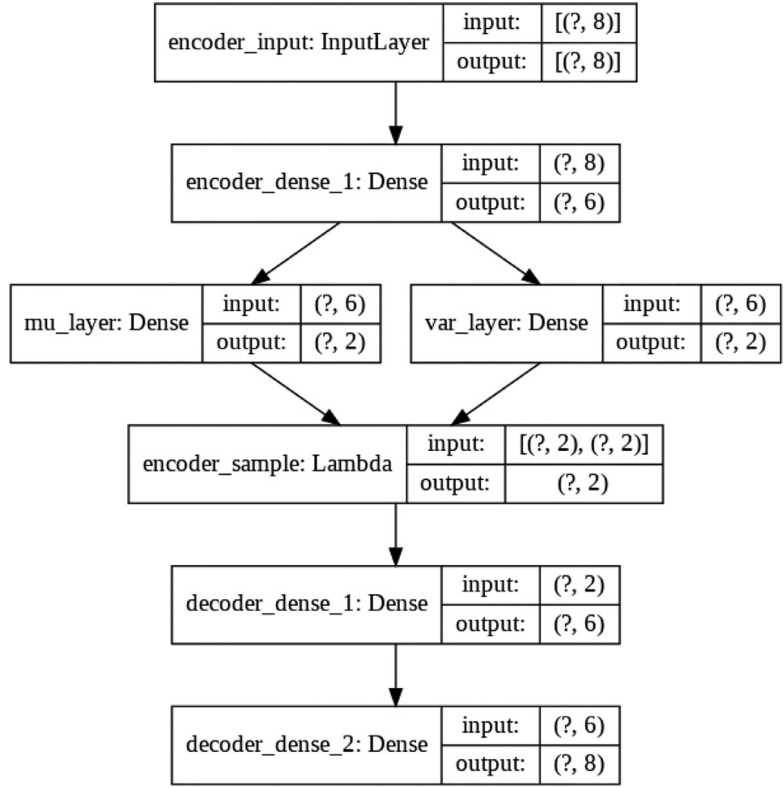

**Fig. 4.** Our VAE to generate more data of the less representative class.

Over-Sampling Technique (SMOTE) [2], Generative Adversarial Network (GAN) [7] and Variational Autoencoders (VAEs) [14] are some examples of oversampling methods, which are also known as generative methods. In this paper, Variational Autoencoders (VAEs) are used to deal with unbalanced dataset because it has demonstrated that it works better than the other known techniques [6].

VAE are part of a group of deep learning techniques known as autoencoders which tries to learn a deep representation of the data by compressing the features. To do so, a symmetric architecture is build in which the number of input neurons is the same as the number of output neurons, having a bottle neck in some hidden layer. When it comes to getting the input itself out of the network, the network is able to learn a deep and compressed representation of the data in the bottle neck layer. All of the layers before the bottle neck one is usually called an encoder, and all the layers after it, makes up the decoder.

The main feature of VAE is that the encoder learns a normal distribution of the data instead of just representing each sample. With this distribution, the latent layer (bottle neck layer) samples a new element with the newly learnt distribution. Once the VAE is trained, each time a new element of the dataset is introduced into the net, it will generate a new one which fits into the same normal distribution of the data. With this approximation, synthetic data which is similar to the original one can be generated.

An example of an autoencoder and a variational autoencoder are shown in the Fig. 1 using an example image.

The vanilla autoencoder is trained using a mean squared error loss over the original element and the reconstructed one. However, in VAE we have to add a new part to the loss function which fixes latent space distribution close to a normal distribution using the Kulback–Leibler divergence (see Eq. (6)).

$$VAE\_LOSS = ||x - \bar{x}||^2 + KL[N(\mu_x, \sigma_x), N(0, 1)] \quad (6)$$

where $\bar{x}$ is the reconstruction of $x$, and $N(\mu_x, \sigma_x)$ a normal distribution with mean $\mu_x$ and standard deviation $\sigma_x$. $KL[p, q]$ is the Kulback–Leilber divergence defined in Eq. (7)

$$KL[p, q] = -\int p(x) \log q(x) dx + \int p(x) \log p(x) dx \quad (7)$$

### 2.3. Feature augmentation: sparse autoencoder (SAE)

Sparse autoencoders (SAEs) are a kind of autoencoder but with the peculiarity that the latent space layer has more neurons than the input and the output. However, an L1 regularization term was added to this latent space layer to force the network to just use some of its neurons each time. Eq. (8) shows the L1 norm.

$$||\mathbf{W}||_1 = |w_1| + |w_2| + \ldots + |w_N| \quad (8)$$

where $w_x$ is the weight of the connection $x$ and $N$ is the number of connection in the layer. L1 regularization adds $\mathbf{W}$ to the loss function forcing the network to have small weights (see Eq. (9))

$$Loss = Error(y, \hat{y}) + \lambda \sum_{i=1}^{N} |w_i| \quad (9)$$

where $y$ is the label of the sample, $\hat{y}$ is the predicted value and $\lambda$ is the multiplication factor of the regularization term. As bigger lambda is, more influence of the regularization in the whole loss calculation. With this, the network learns to represent our initial data with more features, allowing us to analyze the data from another perspective.

In Fig. 2 the typical architecture of a SAE can be seen.

### 2.4. Data classification: multilayer perceptron (MLP)

In order to classify the data, one of the most typical neural network architecture is the multilayer perceptron (MLP) [24]. This network is made up of one input layer, one output layer and one or more hidden layers. In deep learning, more than one layer is usually used in order to learn complex information on the input data.

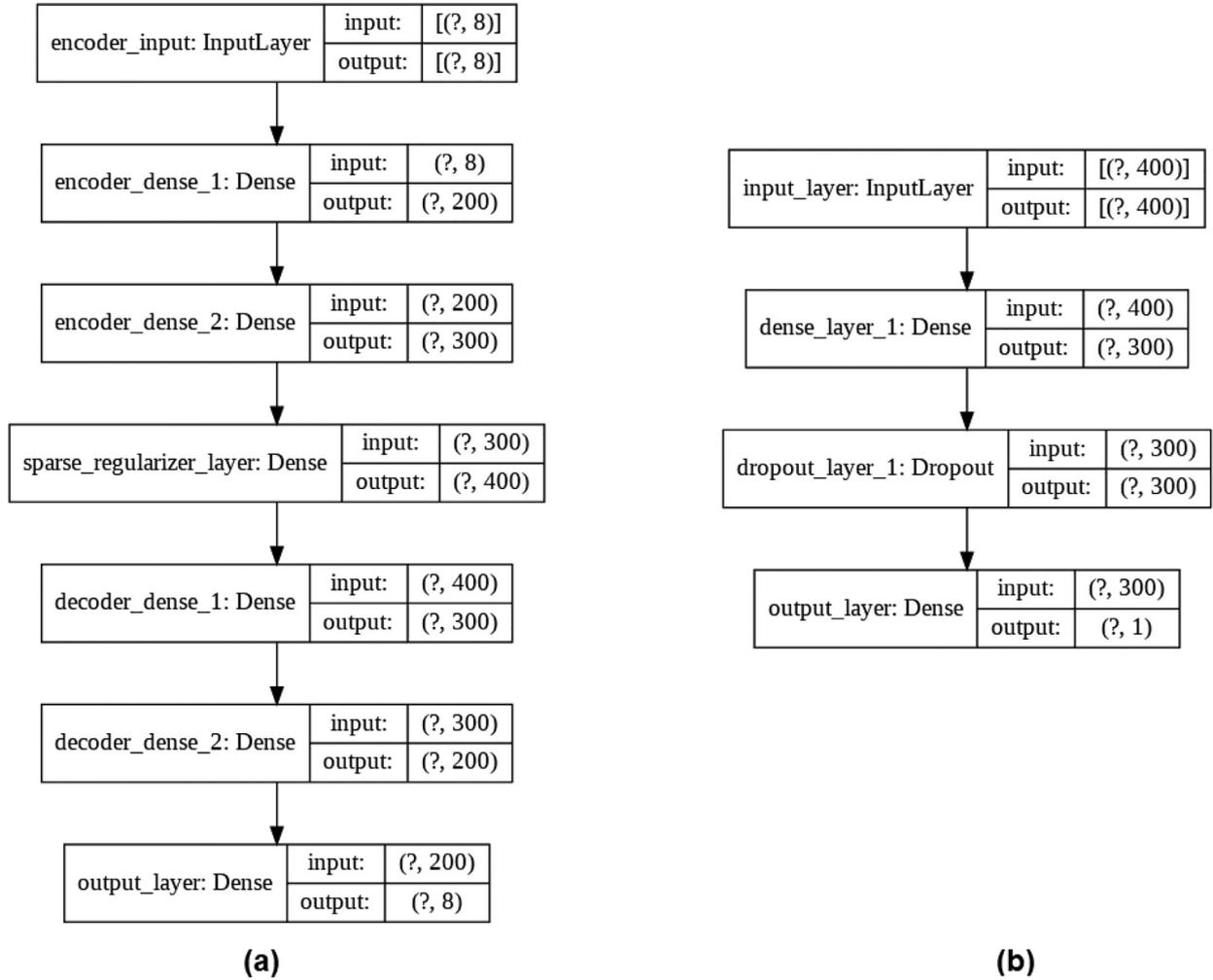

**Fig. 5.** (a) Proposed SAE in this research to generate 400 feature data. (b) Our MLP scheme.

In an MLP, each neuron of the $L$ layer is fully connected with all the neurons of the $L+1$ layer. When the classification problem is binary, the most common approach uses one output neuron with a sigmoid activation function which represents the probability of the input belonging to the positive class(see Eq. (10)).

$$f(x) = \frac{1}{1 + e^{-x}} \quad (10)$$

*2.5. Data classification: convolutional neural networks (CNN)*

A convolutional neural network (CNN) [11], is a deep learning neural network algorithm which can take in a bi-dimensional input and be able to extract complex features of the data. To do so, in the training process of a CNN classification, the network adjusts the weights of filters in order to carrying out an accurate feature map of each class. A basic modelling of a CNN is represented in Fig. 3

After a convolutional layer, it is common to add a pooling layer. These kinds of layers are used to decrease the number of parameters in the network. This reduces the computational cost and controls overfitting. The most frequent type of pooling is Max-pooling, which takes the maximum value in each window. In order to carry out a classification or a regression problem with the features generated by the convolutional layers, it is necessary to add dense layers at the end of the network.

*2.6. Classical machine learning techniques*

A wide number of well-known machine learning techniques have also been used to compare the results with those obtained in our experiments:

**Decision Tree** is a model in which each internal node (not leaf) is tagged with an input characteristic. Arcs that come from a node tagged with an input feature are tagged with each of the possible values of the target or output feature, or the arc leads to a subordinate decision node on a different input feature. Each leaf in the tree is labeled with a class or probability distribution over the classes, which means that the dataset has been classified by the tree into a specific class or into a particular probability distribution. We have carried out a grid search to estimate the best value for the max_leaf_nodes and the criterion.

**Random Forest** is a combination of decision trees such that each tree depends on the values of a random vector tested independently and with the same distribution for each of them. We have carried out a grid search to estimate the best value for the max_leaf_nodes, the number of trees and the criterion.

**SVM** converts the training samples to points in hyperspace to maximize the width of the gap between the decision boundary of the two categories. Also, using kernels, SVM can learn non-linear separations of the data. We have carried out a grid search to estimate the best value for kernel type, the decision function shape and the nu parameter.

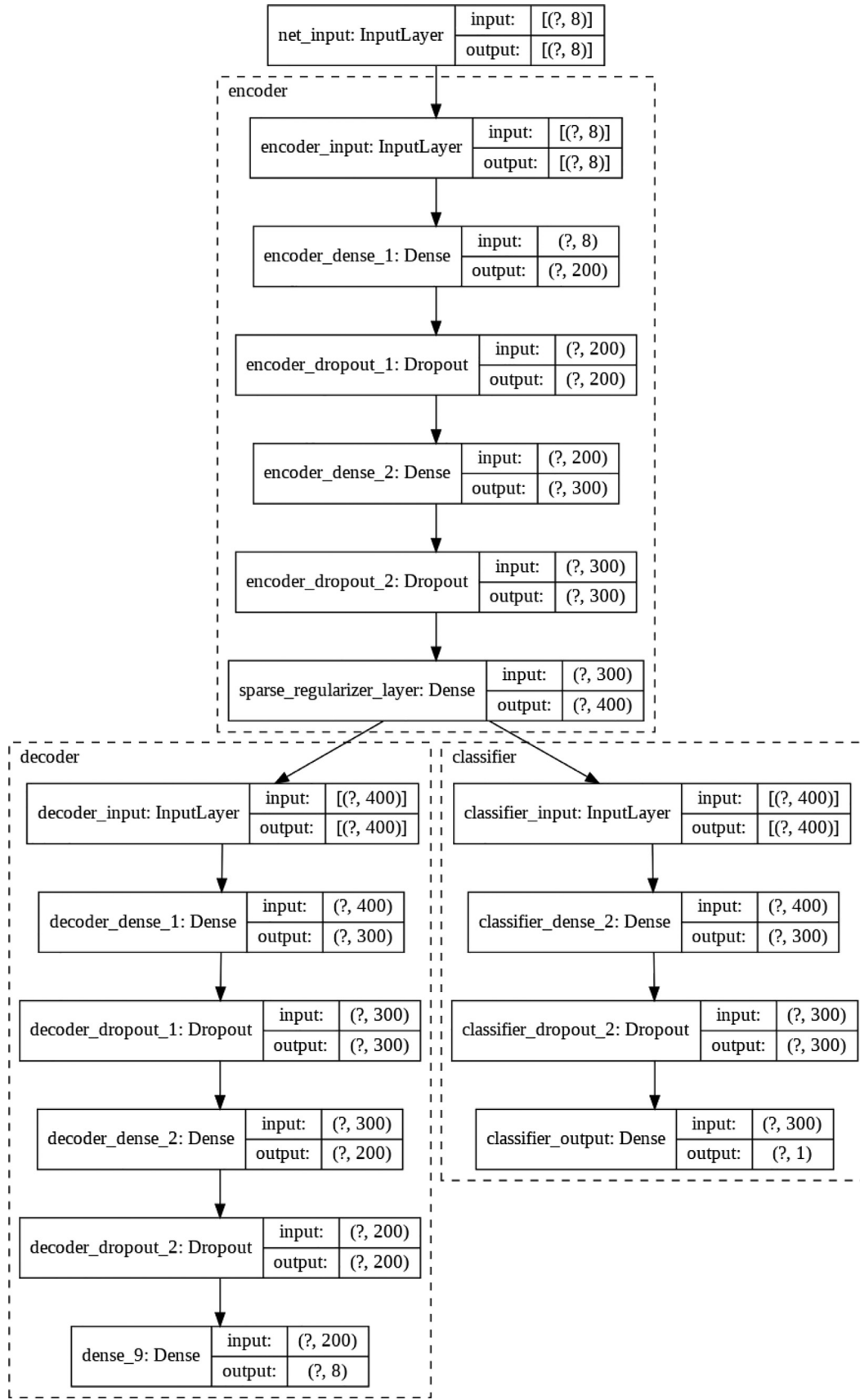

**Fig. 6.** Autoencoder with latent space classifier.

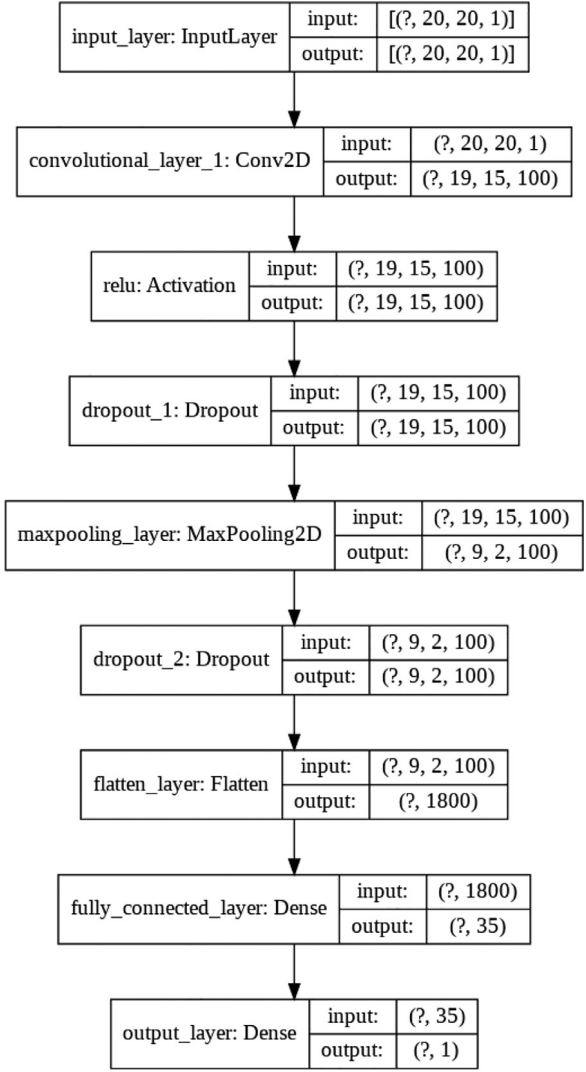

**Fig. 7.** Proposed convolutional neural network architecture.

**k-NN** is a nonparametric classification method, which estimates the value of the probability density function that an element belongs to a given class from the information provided by the set of training elements that are closest to it. We have carried out a grid search to estimate the best value for the number of neigbors and the type of algorithm used (ball tree, kd tree or brute).

**XGBoost** is a decision tree-based ensemble machine learning algorithm that uses a gradient boost framework designed to minimize execution speed and maximize performance. We have carried out a grid search to estimate the best value for the minimum sum of instance weight (hessian) needed in a child, the gamma parameter, the subsample ratio of the training instances, the subsample ratio of columns when constructing each tree and the maximum depth of a tree.

## 3. Experiments and results

### 3.1. Dataset

The Pima Indians Diabetes Database (PIDD) is sourced from the UCI machine learning repository [13]. It is made up of 768 samples, each one with eight features: Pregnancies, Glucose, BloodPressure, SkinThickness, Insulin, BMI, DiabetesPedigreeFunction and Age. Five hundred samples are labeled as non diabetic (class 0) and the rest, two hundred and sixty eight samples, belong to the diabetic class (class 1).

### 3.2. Experimental setup

First of all, for training process in all the models proposed, we have split our dataset for training (90%) and testing (10%) to avoid misleading results, showing the results of the test subset.

The first step of the present work was the normalisation of the data. Min-Max feature normalization using training subset has been carried out to set the values of the numeric columns in the dataset to a common scale [0,1], without distorting differences in the ranges of values. Test subset has been also normalized using the training parameters.

Furthermore, the pregnancies feature was transformed into 1 or 0 (pregnancy or not respectively) instead of representing the number of pregnancies. In the original dataset, some features are 0, but this value must be considered as a missing value because, according to the experts, features such as glucose, blood pressure, insulin, etc. cannot be 0. We solved this problem by filling in the missing values with the mean value of their column in the training subset.

After that, the class distribution of the dataset was evaluated by taking two different values into account: Diabetic represented by 1 and non diabetic represented by 0. The training split contains 449 class 0 elements and 242 belonging to class 1. Although the data is not quite unbalanced, a Variational Autoencoder (VAE) was trained over this subset in order to generate more data on the less representative class. The scheme of our VAE is detailed in Fig. 4.

#### 3.2.1. Sparse autoencoder and multi layer perceptron classifier trained separately (SAE + MLP)

After this process, the training split is made up of 449 elements of class 0 and 484 of class 1. Deep learning techniques require large amounts of data and features in order to improve their learning process. We proposed the used of a Sparse Autoencoder (SAE) to transform our eigth-element data to 400-element data in order to improve the performance of the classification. The process was carried out using the Sparse Autoencoder represented in Fig. 5(a).

After all these steps, our data is made up of 1036 elements described by 400 features each. This dataset was classified using a multilayer perceptron. The Multi Layer Perceptron (MLP) includes two dropout layers to avoid overfitting and it was training for more than 400 epochs. The architecture of our MLP is shown in Fig. 5(b).

#### 3.2.2. Sparse autoencoder and multi layer perceptron classifier trained jointly (SAE with MLP)

It has been decided to train an autoencoder with classifier architecture all together. With this, the features learnt in the latent space during the autoencoder training are also influenced by the class of the samples thanks to the classifier. This architecture was trained for more than 400 epochs using the Adam optimizer. Fig. 6 shows the diagram of the layers of this architecture.

**Table 1**
Comparison with state of the art methods.

| Method | Accuracy | Authors |
|---|---|---|
| Hierarchical Fuzzy Classification | 79.71 | Feng et al [4] |
| NSGA-II-Stacking | 83.80 | Singh and Singh [26] |
| Re-RX with J48graft | 83.83 | Hayashi et al [8] |
| RLEFRBS | 84.00 | Mansourypoor and Asadi [20] |
| Modified Artificial Bee Colony | 84.21 | Beloufa and Chikh [1] |
| ANN + FNN | 84.24 | Kahramanli and Allahverdi [9] |
| RST-BatMiner | 85.33 | Cheruku et al [3] |
| Stacked autoencoders | 86.26 | Kannadasan et al [10] |
| TLBO-FWNN | 88.67 | Majeed-Alneamy et al [19] |
| Fuzzy SVM | 89.02 | Lukmanto et al [17] |
| PCA + ANFIS | 89.47 | Polat and Güneş [23] |
| SAE with CNN | 92.31 | Ours |

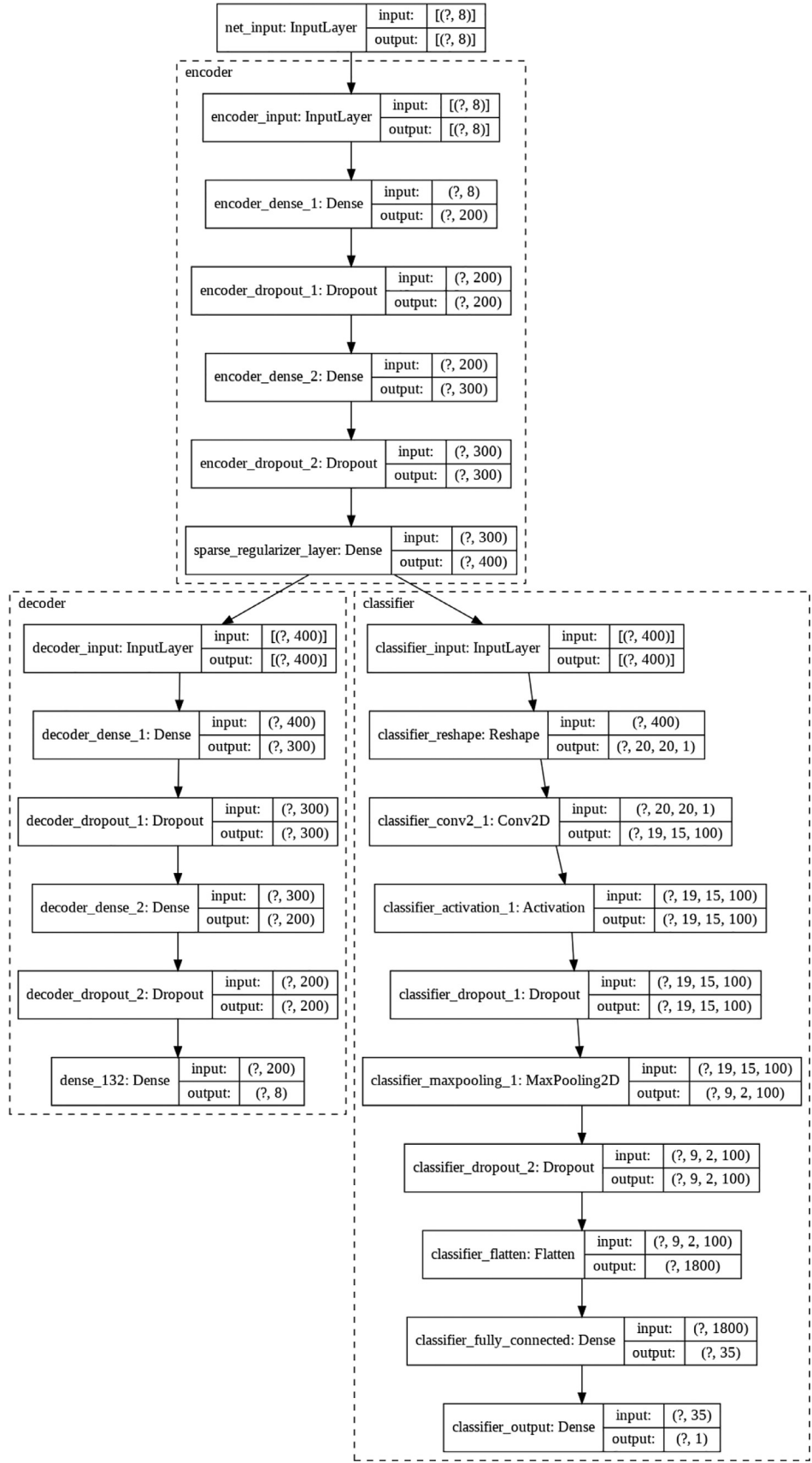

**Fig. 8.** Proposed autoencoder with a classification inline.

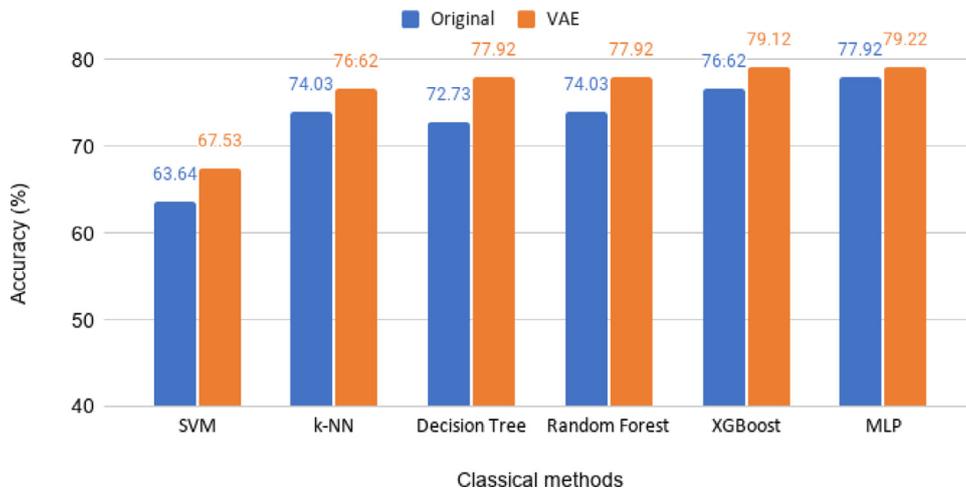

**Fig. 9.** Comparison of classical methods with both original and oversampled datasets.

### 3.2.3. Sparse autoencoder and convolutional neural network classifier trained separately (SAE + CNN)

A convolutional neural network was trained for our third proposal. Eight features for each sample are not sufficient to train a convolutional neural network (CNN), so in this experiment, we proposed the use of an Sparse Autoencoder (SAE) to transform our eight-element data to 400 features in the same way as in previous sections and then, these 400 elements were transformed to a $20 \times 20$ matrix data. As we can see in Fig. 7, our CNN architecture includes two dropout layers (with dropout rate = 0.2) to deal with the overfitting problem and one maxpooling layer to reduce the dimensionality. The convolutional layer has 100 filters with a kernel size of (2,6) and stride of 1. Maxpooling layer has a pool size of (2,6) too. All of these hyperparameters were chosen after a grid search evaluation. CNN has been trained for more than 600 epochs and 50 elements as batch size.

### 3.2.4. Sparse autoencoder and CNN classifier trained jointly (SAE with CNN)

The CNN classifier has also been combined with the jointly proposed autoencoder. The CNN classifier is exactly the same as that used in the previous section but the CNN classification is carried out inside the autoencoder. This architecture was trained for more than 600 epochs using Adam optimizer. Fig. 8 shows the architecture diagram.

### 3.3. Results

First, a comparison was made between the original dataset and the oversampled VAE dataset using classical machine learning techniques. Fig. 9 shows the results obtained after an exahustive hyperparameter grid search. As we can see, every model trained with VAE dataset outperform the trained model with the original dataset. The best result has been achieved by using MLP with a 79.22% of accuracy in the test subset.

As the PIMA Indians dataset only contains 8 features, a feature augmentation has been carried out using the Sparse autoencoder to extract 400 new features. Thanks to the high number of features we have extracted, a convolutional neural network can be trained by reshaping the 400 features into a 2D array of $20 \times 20$. Also a MLP has been trained with the new SAE features as it was the model with the best performance in the previous experiment (see Fig. 9).

Moreover, a jointly net which combines SAE and the classifier (MLP or CNN) has been implemented in order to increase the feature extraction ability by taking into account the classifier information obtained as feedback in the backpropagation algorithm.

Fig. 10 shows the results obtained with the jointly nets (SAE with CNN and SAE with MLP), with the classification after the SAE

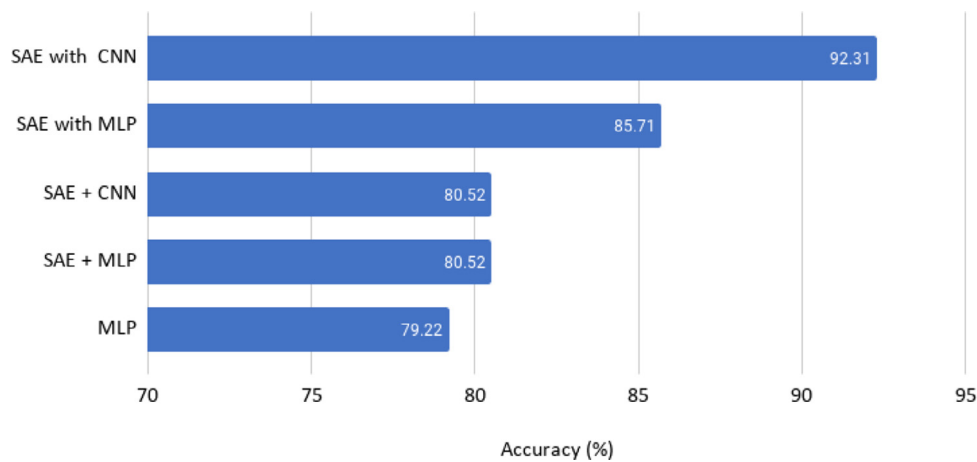

**Fig. 10.** Results achieved using SAE with a network, a network after SAE and a MLP without SAE.

feature augmentation (SAE + CNN and SAE + MLP) and the previous result achieved without SAE (MLP).

The best performance was achieved using SAE with CNN. As we can see, training the sparse autoencoder with the classifier all in the same architecture improves the results in comparison with training them sequentially. It is important to notice how MLP and CNN obtains the same results when SAE is trained before the classification training. However, when the SAE is trained jointly, CNN outperforms MLP in a 7.7% of accuracy. It indicates that CNN interfers in the SAE feature extraction by forcing it to extract more relevant features with spatial location information. MLP also modify the feature extraction carried out by SAE but the improvement is clearly worse than in the convolutional net.

A one-way between subjects analysis of variance (ANOVA) was performed to compare the effect of applying different neural networks to the same data on precision. There was a significant effect of neural network applied on accuracy at the $p < 0.05$ level for the five conditions $[F(4, 51) = 574.929, \ p < 0.001]$. Post hoc comparisons using the Tukey HSD test indicated that the mean score for the SAE with CNN experiment (M = 92.31, SD = 1.04) was significantly different than SAE with MLP (M = 85.71, SD = 0.66), SAE + CNN (M = 80.52, SD = 0.65), SAE+MLP (M = 80.52, SD = 0.65), MLP (M = 79.22, SD = 0.77). Taken together, these results suggest that SAE with CNN experiment really do have an effect on accuracy. Specifically, our results suggest that when SAE with CNN is applied, accuracy is significantly higher.

Furthermore, we have evaluated our results with the most recent papers on the state of the art with this dataset. Since all other papers do not specify the concrete data split for training and testing, the results can not be compared exactly. However, every paper show their best result demonstrating that with this dataset our proposal outperforms the state of the art. In Table 1 we can see the comparison.

The SAE with CNN proposal obtains the greatest precision, surpassing all methods seen in the state of the art.

## 4. Conclusions

This paper proposes methods based on in-depth learning combined with augmentation techniques to address the prediction of diabetes using a popular data set called Pima Indian Diabetes. This dataset is made up of 768 examples with just 8 features per sample and a unbalanced number of classes. In the preprocessing step, the data set has been augmented using a Variational Autoencoder (VAE) and the number of features has been expanded with a Sparse Autoencoder. Thanks to this, it was possible to train a convolutional neural network to carry out the classification step. A new architecture approach which combines the Sparse Autoencoder and the Convolutional Classifier was proposed obtaining a 92.31% of accuracy, outperforming all the other techniques shown in the state of the art.

Using a multi task neural network with SAE and CNN jointly, the contribution is not only the better classification of the diabetes samples, but also the way to generate new features for the dataset. The proposed architecture can be used in a wide new research fields. It helps CNN to deal with structured data by rearranging their data using SAE adding the optimum spatial representation by reordering the features and creating new ones as combination of them

Using a full deep learning pipeline for data preprocessing and classification has demonstrate to be very promising in the diabetes detection field outperforming the state-of-the-art proposals.

Although these results are very promising, this work is limited to the small number of samples in the dataset studied. It is very possible that the results can be improved by creating, as future work, a new dataset with more valuable characteristics and more individuals that allow a better generalization of the learning model.

## Declaration of Competing Interest

Authors declare that they have no conflict of interest.

## CRediT authorship contribution statement

**María Teresa García-Ordás:** Conceptualization, Formal analysis, Investigation, Methodology, Software, Validation, Writing - original draft. **Carmen Benavides:** Investigation, Validation, Writing - review & editing. **José Alberto Benítez-Andrades:** Conceptualization, Methodology, Software, Writing - review & editing. **Héctor Alaiz-Moretón:** Methodology, Writing - review & editing. **Isaías García-Rodríguez:** Formal analysis, Investigation, Validation, Writing - review & editing.

## Acknowledgements

We gratefully acknowledge the support provided by the Consejería de Educación, Junta de Castilla y León throught project LE078G18. UXXI2018/000149. U-220.

## References

[1] F. Beloufa, M.A. Chikh, Design of fuzzy classifier for diabetes disease using modified artificial Bee Colony algorithm, Comput. Methods Prog. Biomed. 112 (1) (2013) 92–103, doi:10.1016/j.cmpb.2013.07.009.
[2] N.V. Chawla, K.W. Bowyer, L.O. Hall, W.P. Kegelmeyer, SMOTE: synthetic minority over-sampling technique, J. Artif. Intell. Res. 16 (2002) 321–357, doi:10.1613/jair.953.
[3] R. Cheruku, D.R. Edla, V. Kuppili, R. Dharavath, RST-BatMiner: a fuzzy rule miner integrating rough set feature selection and Bat optimization for detection of diabetes disease, Appl. Soft Comput. 67 (2018) 764–780, doi:10.1016/J.ASOC.2017.06.032.
[4] T.C. Feng, T.H.S. Li, P.H. Kuo, Variable coded hierarchical fuzzy classification model using DNA coding and evolutionary programming, Appl. Math. Model. 39 (23-24) (2013) 7401–7419, doi:10.1016/j.apm.2015.03.004.
[5] S.S. G., M. K., Diagnosis of diabetes diseases using optimized fuzzy rule set by grey wolf optimization, Pattern Recognit. Lett. 125 (2019) 432–438, doi:10.1016/J.PATREC.2019.06.005.
[6] M.T. García-Ordás, J.A. Benítez-Andrades, I. García-Rodríguez, C. Benavides, H. Alaiz-Moretón, Detecting respiratory pathologies using convolutional neural networks and variational autoencoders for unbalancing data, Sensors 20 (4) (2020) 1214, doi:10.3390/s20041214.
[7] I.J. Goodfellow, J. Pouget-Abadie, M. Mirza, B. Xu, D. Warde-Farley, S. Ozair, A. Courville, Y. Bengio, Generative adversarial nets, in: Advances in Neural Information Processing Systems, 3, Neural Information Processing Systems Foundation, 2014, pp. 2672–2680, doi:10.3156/jsoft.29.5_177_2.
[8] Y. Hayashi, S. Yukita, Rule extraction using recursive-rule extraction algorithm with J48graft combined with sampling selection techniques for the diagnosis of type 2 diabetes mellitus in the Pima Indian dataset, Inform. Med. Unlocked 2 (2016) 92–104, doi:10.1016/J.IMU.2016.02.001.
[9] H. Kahramanli, N. Allahverdi, Design of a hybrid system for the diabetes and heart diseases, Expert Syst. Appl. 35 (1-2) (2008) 82–89, doi:10.1016/J.ESWA.2007.06.004.
[10] K. Kannadasan, D.R. Edla, V. Kuppili, Type 2 diabetes data classification using stacked autoencoders in deep neural networks, Clin. Epidemiol. Glob. Health 7 (4) (2019) 530–535, doi:10.1016/j.cegh.2018.12.004.
[11] A. Karpathy, G. Toderici, S. Shetty, T. Leung, R. Sukthankar, L. Fei-Fei, Large-scale video classification with convolutional neural networks, in: 2014 IEEE Conference on Computer Vision and Pattern Recognition, 2014, pp. 1725–1732, doi:10.1109/CVPR.2014.223.
[12] H. Kaur, V. Kumari, Predictive modelling and analytics for diabetes using a machine learning approach, Appl. Comput. Inform. (2018), doi:10.1016/J.ACI.2018.12.004.
[13] K. Kayaer, T. Yildirim, Medical Diagnosis on Pima Indian Diabetes Using General Regression Neural Networks, Technical Report (2003).
[14] D.P. Kingma, M. Welling, Stochastic Gradient VB and the Variational Auto-Encoder, Technical Report (2013).
[15] A. Lonappan, G. Bindu, V. Thomas, J. Jacob, C. Rajasekaran, K.T. Mathew, Diagnosis of diabetes mellitus using microwaves, J. Electromagn. Waves Appl. 21 (10) (2007) 1393–1401, doi:10.1163/156939307783239429.
[16] I.B. Ltaifa, L. Hlaoua, L.B. Romdhane, Hybrid deep neural network-based text representation model to improve microblog retrieval, Cybern. Syst. 51 (2) (2020) 115–139, doi:10.1080/01969722.2019.1705548.

[17] R.B. Lukmanto, Suharjito, A. Nugroho, H. Akbar, Early detection of diabetes mellitus using feature selection and fuzzy support vector machine, Procedia Comput. Sci. 157 (2019) 46–54, doi:10.1016/J.PROCS.2019.08.140.

[18] T. Mahboob Alam, M.A. Iqbal, Y. Ali, A. Wahab, S. Ijaz, T. Imtiaz Baig, A. Hussain, M.A. Malik, M.M. Raza, S. Ibrar, Z. Abbas, A model for early prediction of diabetes, Inform. Med. Unlocked 16 (2019) 100204, doi:10.1016/J.IMU.2019.100204.

[19] J.S. Majeed Alneamy, Z. A. Hameed Alnaish, S. Mohd Hashim, R.A. Hamed Alnaish, Utilizing hybrid functional fuzzy wavelet neural networks with a teaching learning-based optimization algorithm for medical disease diagnosis, Comput. Biol. Med. 112 (2019) 103348, doi:10.1016/J.COMPBIOMED.2019.103348.

[20] F. Mansourypoor, S. Asadi, Development of a reinforcement learning-based evolutionary fuzzy rule-based system for diabetes diagnosis, Comput. Biol. Med. 91 (2017) 337–352, doi:10.1016/J.COMPBIOMED.2017.10.024.

[21] D.M. Nathan, E. Barret-Connor, J.P. Crandall, S.L. Edelstein, R.B. Goldberg, E.S. Horton, W.C. Knowler, K.J. Mather, T.J. Orchard, X. Pi-Sunyer, D. Schade, M. Temprosa, Long-term effects of lifestyle intervention or metformin on diabetes development and microvascular complications: the DPP outcomes study, Lancet Diabetes Endocrinol. 3 (11) (2016) 866–875, doi:10.1016/S2213-8587(15)00291-0.Long-term.

[22] E.A. Olawsky, Y. Zhang, A.C. Alvear, L.E. Eberly, L.S. Chow, 864-P: hyperglycemia drives glycemic variability in patients with Type 2 diabetes (T2DM), Diabetes 69 (Supplement 1) (2020) 864–P, doi:10.2337/db20-864-p.

[23] K. Polat, S. Güne, An expert system approach based on principal component analysis and adaptive neuro-fuzzy inference system to diagnosis of diabetes disease, Digit. Signal Process. 17 (4) (2007) 702–710, doi:10.1016/j.dsp.2006.09.005.

[24] D.E. Rumelhart, G.E. Hinton, R.J. Williams, Learning Internal Representations by Error Propagation, MIT Press, Cambridge, MA, USA, p. 318–362.

[25] K. Sim, J. Yang, W. Lu, X. Gao, MaD-DLS: mean and deviation of deep and local similarity for image quality assessment, IEEE Trans. Multimed. (2020) 1, doi:10.1109/tmm.2020.3037482.

[26] N. Singh, P. Singh, Stacking-based multi-objective evolutionary ensemble framework for prediction of diabetes mellitus, Biocybern. Biomed. Eng. 40 (1) (2020) 1–22, doi:10.1016/J.BBE.2019.10.001.

[27] D. Sisodia, D.S. Sisodia, Prediction of diabetes using classification algorithms, Procedia Comput. Sci. 132 (2018) 1578–1585, doi:10.1016/J.PROCS.2018.05.122.

[28] Y. Wang, Z. Pan, X. Yuan, C. Yang, W. Gui, A novel deep learning based fault diagnosis approach for chemical process with extended deep belief network, ISA Trans. 96 (2020) 457–467, doi:10.1016/j.isatra.2019.07.001.

[29] J. Yang, C. Wang, B. Jiang, H. Song, Q. Meng, Visual perception enabled industry intelligence: state of the art, challenges and prospects, IEEE Trans. Ind. Inform. 17 (3) (2021) 2204–2219, doi:10.1109/TII.2020.2998818.

[30] X. Yuan, L. Li, Y. Shardt, Y. Wang, C. Yang, Deep learning with spatiotemporal attention-based LSTM for industrial soft sensor model development, IEEE Trans. Ind. Electron. (2020) 1, doi:10.1109/TIE.2020.2984443.

[31] X. Yuan, C. Ou, Y. Wang, C. Yang, W. Gui, A layer-wise data augmentation strategy for deep learning networks and its soft sensor application in an industrial hydrocracking process, IEEE Trans. Neural Netw. Learn. Syst. (2019) 1–10, doi:10.1109/TNNLS.2019.2951708.

[32] X. Yuan, J. Zhou, B. Huang, Y. Wang, C. Yang, W. Gui, Hierarchical quality-relevant feature representation for soft sensor modeling: a novel deep learning strategy, IEEE Trans. Ind. Inform. 16 (6) (2020) 3721–3730, doi:10.1109/TII.2019.2938890.

**María Teresa García-Ordás, Ph.D.** was born in León, Spain, in 1988. She received her degree in Computer Science from the University of León in 2010, and her Ph.D. in Intelligent Systems in 2017. She was a recipient of a special mention award for the best doctoral thesis on digital transformation by Tecnalia. Since 2019, she works as teaching assistant at the University of León. Her research interests include computer vision and deep learning. She has published several articles in impact journals and patents. She has participated in many conferences all over the world.

**José Alberto Benítez-Andrades, Ph.D.** was born in Granada, Spain, in 1988. He has received his degree in Computer Science from the University of León, and the Ph.D. degree in Production and Computer Engineering in 2017 (University of Leon). He was part time instructor who kept a parallel job from 2013 to 2018 and since 2018 he works as teaching assistant at the University of Leon. His research interests include artificial intelligence, knowledge engineering, semantic technologies. He was a recipient of award to the Best Doctoral Thesis 2018 by Colegio Profesional de Ingenieros en Informática en Castilla y León in 2018.

**Isaías García-Rodríguez, Ph.D.** received his Bachelor degree in Industrial Technical Engineering from the University of León (Spain) in 1992 and his Master degree in Industrial Engineering from the University of Oviedo (Spain) in 1996. Isaías obtained his Ph.D. in Computer Science from the University of León in 2008, where he is currently a lecturer. His current research interests include practical applications of Software Defined Networks, Network Security and applied Knowledge Engineering techniques. He has published different scientific papers in journals, Conferences and Symposia around the world.

**Carmen Benavides, Ph.D.** received her Bachelor degree in Industrial Technical Engineer from the University of León (Spain) in 1996 and her Master degree in Electronic Engineering from the University of Valladolid (Spain) in 1998. Carmen obtained her Ph.D. in Computer Science from the University of León in 2009 and she works as an Assistant Professor at the same University since 2001. Her research interests are focused on applied Knowledge Engineering techniques, practical applications of Software Defined Networks and Network Security. She has organized several congresses, and has presented and published different papers in Journals, Conferences and Symposia.

**Héctor Alaiz-Moretón, Ph.D.** received his degree in Computer Science, performing the final project at Dublin Institute of Technology, in 2003. He received his Ph.D. in Information Technologies in 2008 (University of Leon). He has worked like a lecturer since 2005 at the School of Engineering at the University of Leon. His research interests include knowledge engineering, machine & deep learning, networks communication and security. He has several works published in international conferences, as well as books and scientific papers in peer review journals. He has been member of scientific committees in conferences. He has headed several Ph.D. Thesis and research projects.